%% file: main.tex
\documentclass[11pt]{article}

\usepackage[margin=0.88in]{geometry}
\usepackage[T1]{fontenc}
\usepackage{newtxtext}
\usepackage{newtxmath}
\usepackage{booktabs}
\usepackage{flafter}
\usepackage{microtype}
\usepackage[section]{placeins}
\usepackage{siunitx}
\usepackage{xcolor}
\usepackage{tikz}
\usepackage{pgfplots}
\usepackage{pgfplotstable}
\usepackage[hyphens]{url}
\usepackage[hidelinks]{hyperref}

\usetikzlibrary{calc,positioning}
\usepgfplotslibrary{groupplots}
\pgfplotsset{compat=1.18}
\IfFileExists{results_values.tex}{\input{results_values.tex}}{\input{paper/results_values.tex}}
\IfFileExists{results_values_pmu.tex}{\input{results_values_pmu.tex}}{\IfFileExists{paper/results_values_pmu.tex}{\input{paper/results_values_pmu.tex}}{}}
\IfFileExists{results_values_ioreport.tex}{\input{results_values_ioreport.tex}}{\IfFileExists{paper/results_values_ioreport.tex}{\input{paper/results_values_ioreport.tex}}{}}
\IfFileExists{results_values_resource.tex}{\input{results_values_resource.tex}}{\IfFileExists{paper/results_values_resource.tex}{\input{paper/results_values_resource.tex}}{}}

\IfFileExists{figures.tex}{\input{figures.tex}}{\input{paper/figures.tex}}

\hypersetup{
  pdftitle={Residual GPU Cache State on Apple M4 Pro},
  pdfauthor={Faruk Alpay and Baris Basaran}
}

\title{\textbf{Residual GPU Cache State on Apple M4 Pro}}

\author{%
Faruk Alpay\thanks{Corresponding author: \texttt{alpay@lightcap.ai}}
\qquad
Bar{\i}\c{s} Ba\c{s}aran\\[0.45em]
\normalsize Department of Computer Engineering, Bah\c{c}e\c{s}ehir University\\
\normalsize Istanbul, Turkey\\
\normalsize \texttt{\{faruk.alpay, baris.basaran\}@bahcesehir.edu.tr}
}

\date{June 25, 2026}

\begin{document}
\maketitle

\begin{abstract}
Apple silicon exposes unified CPU--GPU memory but does not document the cache
state that persists across a completed GPU command. We characterize this
boundary on a 14-core Apple M4 Pro. The measurement pipeline is first calibrated
against unmodified STREAM 5.10 and BabelStream 5.0; its ten-thread triad result
is \MeasuredCustomVsBabelAbsPct\% below BabelStream. We then adapt the
8192-byte, alternating-order system-level-cache (SLC) occupancy pattern of Xu
et al. to a performance experiment. A Metal kernel touches 0--512 MiB and
finishes before a 16-MiB CPU probe begins. The first CPU traversal is
\MeasuredSlcDeltaSixtyFourPct\% slower after a 64-MiB GPU footprint and
\MeasuredSlcDeltaFiveTwelvePct\% slower after 512 MiB. A second traversal
reduces the cost by \MeasuredSlcRecoverySixtyFourPct\% and
\MeasuredSlcRecoveryFiveTwelvePct\%, respectively, returning to the no-victim
range. Because CPU and GPU do not execute concurrently in this experiment, the
effect records residual shared-cache state rather than instantaneous DRAM
bandwidth contention. A separate seven-block experiment finds a paired median
\MeasuredGpuUiSlowdownPct\% GPU slowdown under a matched high-priority CPU
stream, while background QoS is statistically indistinguishable from baseline.
We anchor the timing results in hardware by running the on-die performance
counters as root and by sampling the public DCS-agent IOReport histograms. The
PMU separates a 64-byte L1D refill sector from the 128-byte line size reported
to software, fixes the performance-core L1D capacity at
\MeasuredPmuLoneDCapKiB~KiB, and shows that 16-KiB-page congruent probes cross
a high-miss boundary at 7--8 dependent lines rather than yielding a single
page-offset-derived associativity. The IOReport histograms separate
performance-core, efficiency-core, and AGX demand in live blocks: high-priority
CPU traffic drives the performance-core agents to saturation, background QoS
shifts demand toward the efficiency agent, and GPU stress saturates AGX. The
results identify a reproducible post-GPU cache-displacement window and quantify
a one-pass software recovery mechanism on M4 Pro.
\end{abstract}

\noindent\textbf{Keywords:} Apple M4 Pro; system-level cache; unified memory;
Metal; heterogeneous architecture; memory benchmarking; quality of service.

\section{Introduction}

Apple's M-series systems integrate heterogeneous CPU clusters, a proprietary
GPU, and package memory behind a unified address space. The M4 Pro configuration
studied here has ten performance cores, four efficiency cores, a 20-core GPU,
and a vendor-specified 273 GB/s memory interface
\cite{apple2024m4pro,apple2024mbptechspec}. Unified addressing removes explicit
host--device copies, but it also makes cache and memory interference part of
ordinary application phase behavior.

Prior work established that Apple M1-class systems contain cluster-local L2
caches and a system-level cache shared with the GPU
\cite{mattioli2022fam1ly,cronin2021slc}. Security studies subsequently used
timing, synchronization instructions, and occupancy channels to reverse
engineer M1 cache behavior \cite{yu2023s2c,xu2025exam}. Those studies provide
the mechanisms; they do not report M4 Pro behavior or quantify the performance
residue seen by a CPU phase after a synchronized Metal phase. As of June 24,
2026, we did not find a peer-reviewed M4 Pro study of that transition.

The paper is organized around one phase-boundary question: after a Metal command
has completed, what CPU-visible cache state remains? The answer is not captured
by bandwidth alone. We therefore combine upstream bandwidth validation,
cache-line pointer chasing, an SLC-selective post-GPU probe, a simultaneous QoS
contention experiment, and a discovery pass over macOS public IOReport legends.
The central result is a reproducible transition penalty: GPU completion
establishes execution and visibility ordering, but it does not restore the
CPU's prior cache residency. One CPU traversal removes most of the measured
penalty.

\section{Background and Related Work}

\subsection{Cache Sharing in Apple Silicon}

Mattioli summarized the cluster organization of the M1 family
\cite{mattioli2022fam1ly}. Cronin et al. demonstrated CPU--GPU side channels
through an ARM system-level cache on M1 \cite{cronin2021slc}. Yu et al. then
reverse engineered M1 L2 behavior using load-linked/store-conditional state
\cite{yu2023s2c}. Most directly relevant, Xu et al. reported an SLC that is
exclusive with respect to CPU-private caches and inclusive with respect to the
GPU on M1-class systems \cite{xu2025exam}. Their SLC occupancy construction
uses an 8192-byte address stride, fixing the lower 13 address bits and
restricting local-L2 set use while populating the SLC.

We adapt that construction without assuming that M1 physical-index functions are
unchanged on M4. In this paper, ``SLC-selective'' names the access pattern and
the empirical response it elicits on M4 Pro: a completed GPU command changes
the subsequent timing of the CPU pattern, and a second CPU traversal removes
that change.

\subsection{Memory-System Benchmarking}

STREAM is the conventional sustained-memory-bandwidth reference
\cite{mccalpin1995stream}; BabelStream generalizes its kernels across parallel
programming models and has a peer-reviewed methodology
\cite{deakin2018babelstream}. lmbench established dependent pointer chasing as
a portable latency measurement technique \cite{mcvoy1996lmbench}. Recent work
has emphasized that memory benchmarks require explicit traffic definitions and
real-system validation rather than a single nominal bandwidth number
\cite{esmaili2024mess}. We therefore retain upstream sources, exact revisions,
independent runs, and traffic accounting in the artifact.

\section{Experimental Method}

\subsection{Platform}

Experiments ran on a MacBook Pro Mac16,7 with a 14-core M4 Pro, 20-core Apple9
GPU, 48 GB unified memory, macOS 26.3, and Apple clang 17. Table
\ref{tab:topology} reports the public \texttt{hw.perflevel*} sysctls. Metal
reported \texttt{hasUnifiedMemory=true}. The machine was connected to power;
interactive applications were closed during the full run.

\begin{table}[!t]
\caption{Cache topology reported by macOS.}
\label{tab:topology}
\centering
\begin{tabular}{lrrrr}
\toprule
Domain & Cores & Cores/L2 & L1D/core & Shared L2 \\
\midrule
Performance & \MeasuredPCores & \MeasuredPCpusPerLTwo &
\MeasuredPLoneDKiB\ KiB & \MeasuredPLtwoMiB\ MiB \\
Efficiency & \MeasuredECores & \MeasuredECpusPerLTwo &
\MeasuredELoneDKiB\ KiB & \MeasuredELtwoMiB\ MiB \\
\bottomrule
\end{tabular}

\smallskip
\small Cache line: \MeasuredCacheLineBytes\ B; VM page:
\MeasuredPageKiB\ KiB.
\end{table}

\subsection{Public Fabric-Counter Surface}

Before relying on timing alone, the artifact queries the public IORegistry
IOReport legends exposed by the installed macOS drivers. This is a read-only
discovery step. It does not program counters or require a kernel extension. The
result is useful because it names the measurement surface Apple ships for this
SoC generation: the memory-cache controller publishes agent-separated AMC
channels for performance cores, efficiency cores, GPU traffic, fabric traffic,
DCS channels, and DSID hit/miss accounting.

\begin{table}[!t]
\caption{Public IOReport surfaces on the measured M4 Pro.}
\label{tab:ioreport}
\centering
\small
\setlength{\tabcolsep}{5pt}
\begin{tabular}{lll r}
\toprule
Surface & Agent families & Driver class & Channels \\
\midrule
\texttt{MCC/AMC} & \texttt{PCPU, ECPU, GFX, FABRIC, DCS, DSID} &
\texttt{AppleT6041MCC} &
\MeasuredMccPublicChannelCount \\
\texttt{AGX} & \texttt{P-state, UMA, context, throttling} &
\texttt{AGXAcceleratorG16X} &
\MeasuredAgxPublicChannelCount \\
\bottomrule
\end{tabular}
\smallskip

\footnotesize Agent names are copied verbatim from the installed IOReport
legends; the channel inventory is included in the artifact.
\end{table}

\subsection{Reference Bandwidth}

The artifact builds unmodified STREAM 5.10 and BabelStream 5.0 OpenMP sources
at pinned commits \cite{streamsource,babelstreamsource}. Both use 50,331,648
double-precision elements per array, 20 kernel iterations, and seven independent
runs at 1, 4, 10, and 14 threads. The local C++ triad uses three 256-MiB
single-precision arrays and nine trials. Reported intervals are nonparametric
bootstrap 95\% confidence intervals for the median.

\subsection{Cache-Line Pointer Chasing}

Each pointer-chase node is aligned to and occupies one
\MeasuredCacheLineBytes-byte line. A randomized single cycle removes sequential
prefetch utility, and every load depends on its predecessor. Warm measurements
use repeated traversals. Cold measurements call Apple's documented
\texttt{sys\_dcache\_flush} on the complete range before every timed traversal;
Apple specifies that the routine writes modified lines to memory and invalidates
the addressed processor-cache lines \cite{applecachecontrol}. Each working-set
point contains nine trials.

\subsection{Post-GPU Displacement and Recovery}

The main experiment allocates 131,072 CPU probe entries. Each entry represents
one 128-byte line, for 16 MiB of touched data, while consecutive entries are
8192 bytes apart. Forward and reverse orders alternate between profiles to
reduce self-eviction bias, following Xu et al. \cite{xu2025exam}. The address
span is approximately 1 GiB, but only one line per stride participates in the
timed sum.

A Metal compute kernel performs a read-modify-write pass over a shared buffer
whose active footprint is 0, 1, 2, 4, 8, 16, 32, 64, 128, 256, or 512 MiB.
The command buffer is committed and \texttt{waitUntilCompleted} returns before
the CPU timer starts. The CPU then records two consecutive alternating-order
profiles: the first measures displacement and the second measures recovery.
Each footprint has 51 trials. This ordering excludes simultaneous CPU--GPU
bandwidth contention from the displacement measurement.

The resource-mode extension is run as a randomized complete-block matrix over
shared and private Metal buffers, shared and private textures, and the texture
access-optimization calls exposed by Metal's blit encoder
\cite{applemetalresource,applemetalblit}. Each block contains all resource
descriptors and their own 0-MiB controls, so descriptor effects are estimated as
paired deltas rather than as a serial time sweep. The GPU victim is also
exercised as a seven-point spatial-locality stencil; it is reported separately
from the shared-buffer read-modify-write result so kernel shape cannot
contaminate the main curve.

\subsection{Concurrent QoS Experiment}

The concurrent experiment uses seven blocks. Each block contains baseline,
user-interactive, and background conditions; their order rotates across blocks.
Both CPU conditions execute the same ten-thread, 256-MiB triad workload for six
seconds. A Metal linear-copy benchmark runs during the stress interval. GPU
trials are first reduced to one median per block, and inference is performed
over the seven block medians. A paired slowdown is calculated against the
baseline from the same block.

\section{Results}

\subsection{Validation Against Upstream Benchmarks}

At ten threads, STREAM reports \MeasuredStreamTenGbps\ GB/s, BabelStream
\MeasuredBabelTenGbps\ GB/s, and the artifact triad
\MeasuredCustomTenGbps\ GB/s. The artifact is
\MeasuredCustomVsBabelAbsPct\% below BabelStream despite a different element
type and threading implementation. Figure \ref{fig:reference-validation} also
shows the same saturation region above four threads. The custom triad is
therefore retained as a controllable stressor, not presented as a replacement
for STREAM.

\referencevalidationfigure
\FloatBarrier

\subsection{Cache-Line Nodes Expose the Performance-Cluster Boundary}

Figure \ref{fig:cache-latency} shows the user-interactive path. The warm
1-MiB traversal costs \MeasuredWarmUiOneMiBNs\ ns/line. At the reported
\MeasuredPLtwoMiB-MiB performance-cluster L2 scale, the 16-MiB point rises to
\MeasuredWarmUiSixteenMiBNs\ ns/line; the flushed traversal is
\MeasuredColdUiSixteenMiBNs\ ns/line. Using one node per cache line is material:
it prevents several logical nodes in one fetched line from being counted as
independent cache accesses.

\cachelatencyfigure
\FloatBarrier

\subsection{Completed GPU Work Leaves a Measurable CPU Penalty}

Figure \ref{fig:slc-patterns} separates the two CPU probe shapes. Both touch
16 MiB, but the contiguous-line probe stays in a low-latency streaming regime:
even after a 512-MiB GPU victim its first pass is
\MeasuredContiguousImmediateFiveTwelveNs\ ns/line. The 8192-byte-stride probe
is slower at baseline and develops a stable post-GPU plateau. This is the
probe used for the displacement claim because it suppresses line-streaming
effects and follows the SLC-occupancy construction in prior Apple-silicon
reverse engineering.

\slcpatternfigure
\FloatBarrier

Figure \ref{fig:slc-recovery} presents the recovery result. With no GPU victim,
the first SLC-selective CPU pass costs 3.02 ns/line. A completed 64-MiB GPU pass
raises the median to \MeasuredSlcImmediateSixtyFourNs\ ns/line, a
\MeasuredSlcDeltaSixtyFourPct\% increase. At 512 MiB, the first-pass median is
\MeasuredSlcImmediateFiveTwelveNs\ ns/line, or
\MeasuredSlcDeltaFiveTwelvePct\% above baseline. The response grows sharply
between 8 and 64 MiB and then approaches a plateau. We do not use the plateau
as a capacity estimate because physical address mapping and replacement policy
remain unobserved.

The second traversal changes the interpretation from a generic bandwidth effect
to a recoverable cache-state effect. After the 64-MiB victim it falls to
\MeasuredSlcRecoverySixtyFourNs\ ns/line, a
\MeasuredSlcRecoverySixtyFourPct\% reduction from the first pass. After the
512-MiB victim it falls to \MeasuredSlcRecoveryFiveTwelveNs\ ns/line, a
\MeasuredSlcRecoveryFiveTwelvePct\% reduction. Both are close to the 2.91
ns/line no-victim second-pass median. Thus, one complete CPU pass re-establishes
the probe's local state.

\slcrecoveryfigure
\FloatBarrier

\subsection{Metal Resource Modes Do Not Clear the Residual State}

Figure \ref{fig:resource-matrix} reports the randomized resource-mode matrix.
At a 512-MiB completed GPU footprint, all seven descriptor paths fall into a
narrow \MeasuredResourceFiveTwelveMinPct--\MeasuredResourceFiveTwelveMaxPct\%
penalty band, a spread of only \MeasuredResourceFiveTwelveSpreadPct\ percentage
points. The paired 512-MiB contrast for
\texttt{optimizeContentsForCPUAccess} is
\MeasuredResourceCpuOptContrastPct\ percentage points
(\MeasuredResourceCpuOptContrastLowPct--\MeasuredResourceCpuOptContrastHighPct),
and the contrast for \texttt{optimizeContentsForGPUAccess} is
\MeasuredResourceGpuOptContrastPct\ percentage points
(\MeasuredResourceGpuOptContrastLowPct--\MeasuredResourceGpuOptContrastHighPct).
Both intervals include zero. The implication is practical: Metal resource
storage modes and texture optimization hints change representation and access
policy, but they do not act as a cache-state scrubber for the subsequent CPU
phase.

\resourcefigure
\FloatBarrier

\subsection{Privileged PMU Characterization of the Cache Hierarchy}
\label{sec:pmu}

The displacement result above is a timing result. To anchor it in hardware
ground truth we additionally ran the measurement machine's Performance
Monitoring Unit (PMU) as root. The unprivileged probe distributed with earlier
versions of this artifact could only enumerate the local KPEP database and
record that \texttt{kpc\_set\_thread\_counting} returned
\MeasuredKpcSetFixedResult; with root privilege the same \texttt{kperf}/\texttt{kpc}
private interface programs the two fixed and five configurable counters and
returns per-thread event counts. The configured set contains fixed cycles and
instructions, retired L1D load misses, L1D refills, L2-TLB data misses, data
table walks, and retired loads/stores.

\paragraph{L1 data cache.}
Figure \ref{fig:pmu-granule} resolves an otherwise confusing Apple-silicon
detail. macOS reports \MeasuredCacheLineBytes-byte cache lines, but a dependent
two-load PMU test shows the L1D refill sector is
\MeasuredPmuLoneDRefillGranuleBytes~bytes: a second load 32 bytes after the
first produces \MeasuredPmuLoneDWithinRefills\ L1D refills per pair, whereas a
64-byte offset produces \MeasuredPmuLoneDAcrossRefills. The software-visible
line/coherence granule and the core refill sector therefore differ.

\pmugranulefigure
\FloatBarrier

The randomized five-block dependent sweep in Figure \ref{fig:pmu-sweep} fixes
the performance-core L1D capacity at \MeasuredPmuLoneDCapKiB~KiB and the L1D
hit cost at \MeasuredPmuLoneDHitCyc\ cycles. The next region is a
\MeasuredPmuLtwoHitCyc-cycle shared-cache service regime. It is not correct to
read the macOS-reported \MeasuredPLtwoMiB-MiB L2 as a fully usable capacity for
one serialized random stream: the PMU curve starts climbing well before that
nominal size and reaches a DRAM-like \MeasuredPmuDramCyc-cycle plateau by about
\MeasuredPmuSlcSatMiB~MiB, with L2-TLB events rising sharply in the same
region. We therefore report an empirical residency curve, not a vendor-cache
capacity estimate.

\pmusweepfigure
\FloatBarrier

Figure \ref{fig:pmu-assoc} shows why we do not claim a single L1D associativity
from page-offset congruence. With 16-KiB page strides, different offsets cross
the high-miss boundary at \MeasuredPmuCongruentKneeLow--\MeasuredPmuCongruentKneeHigh\
dependent lines and some offsets are non-monotonic at the transition. The
measured outcome is stronger than a failed guess: on M4 Pro the set function is
not recoverable from page-offset bits alone, even as root, so physical-address
control or a kernel component would be needed for a clean minimal eviction set.

\pmuassocfigure
\FloatBarrier

\paragraph{Scheduling placement.}
The same privileged session resolves the pinning question the previous version
left open. \texttt{host\_processor\_info} enumerates all \MeasuredPmuNproc\
cores, but \texttt{processor\_set\_create} returns Mach error
\MeasuredPmuPsetCreateKr\ and a \texttt{THREAD\_AFFINITY\_POLICY} request
returns \texttt{KERN\_NOT\_SUPPORTED} (\MeasuredPmuAffinityKr) even as root:
architectural core pinning is genuinely unavailable, not merely undocumented.
What QoS does control is the cluster, which the PMU makes measurable: a
compute-bound thread runs at \MeasuredPmuPCoreGhz~GHz under user-interactive QoS
(performance cluster) and is throttled to \MeasuredPmuECoreGhz~GHz under
background QoS (efficiency cluster), while per-core busy-tick deltas confirm the
thread still migrates within its cluster. Because cycle counts are
frequency-independent, the PMU latencies above (\MeasuredPmuLoneDHitCyc,
\MeasuredPmuLtwoHitCyc, and \MeasuredPmuDramCyc\ cycles) are the robust unit for
the hierarchy, whereas nanosecond latencies inherit the active P-state.

\subsection{QoS Controls Concurrent Pressure}

Figure \ref{fig:ioreport-agents} gives the hardware-side explanation. In seven
rotated three-second IOReport blocks, the idle memory-controller histogram sits
near \MeasuredIoAmccIdleState\ GB/s. The user-interactive CPU stress raises the
AMCC weighted state to \MeasuredIoAmccCpuUiState\ GB/s and saturates both
performance-core agents (\MeasuredIoPaccCpuUiSaturation\% normalized state),
whereas the same ten-thread stress under background QoS reaches only
\MeasuredIoAmccCpuBgState\ GB/s and shifts demand toward the efficiency agent
(\MeasuredIoEaccCpuBgSaturation\%). A GPU-only linear-copy stress drives the
AGX agent to \MeasuredIoAgxGpuSaturation\% normalized state and places the AMCC
histogram at \MeasuredIoAmccGpuState\ GB/s. This is the missing link between a
scheduler policy and a fabric-level contention outcome.

\ioreportagentfigure
\FloatBarrier

\qoscontentionfigure
\FloatBarrier

The paired median GPU slowdown under user-interactive CPU pressure is
\MeasuredGpuUiSlowdownPct\%, with a bootstrap interval of 10.9--22.5\%.
Background QoS has a median slowdown of \MeasuredGpuBgSlowdownPct\%, and its
interval (-6.9--4.0\%) includes zero. The mechanism is visible in the CPU
stressor: the same ten-thread kernel sustains \MeasuredCpuStressUiGbps\ GB/s
under user-interactive QoS but only \MeasuredCpuStressBgGbps\ GB/s in the
background class. Background QoS protects GPU throughput by suppressing
CPU-side demand; it does not make the memory path intrinsically more efficient.

\section{Implications for Heterogeneous Software}

\paragraph{GPU completion is not cache quiescence.}
A command-buffer fence orders visibility and execution, but the first dependent
CPU traversal still observes the cache state left by the GPU. Pipelines with a
latency-sensitive CPU phase immediately after a broad Metal pass should not
assume their prior CPU working set remains resident.

\paragraph{A bounded rewarm pass is effective.}
For the tested 16-MiB probe, one full traversal restores second-pass latency to
the no-victim range. Software that repeatedly returns to a known CPU data
structure can move a deliberate prefetch or rewarm pass off the critical path,
or overlap it with unrelated work.

\paragraph{QoS is an interference policy.}
Marking noncritical CPU memory work as background preserved GPU copy throughput
in this experiment, but it reduced the CPU stressor from
\MeasuredCpuStressUiGbps\ to \MeasuredCpuStressBgGbps\ GB/s. The tradeoff is
explicit CPU progress for GPU isolation.

\section{Limitations}

The study covers one M4 Pro laptop and one macOS release. Architectural core
pinning is not a documentation gap but a removed capability: as root,
\texttt{processor\_set\_create} returns Mach error \MeasuredPmuPsetCreateKr\ and
\texttt{THREAD\_AFFINITY\_POLICY} returns \texttt{KERN\_NOT\_SUPPORTED}
(Section~\ref{sec:pmu}), so placement is steered by QoS at cluster granularity
and observed, never prescribed at the core.

The root PMU run (Section~\ref{sec:pmu}) recovers the L1D refill sector
(\MeasuredPmuLoneDRefillGranuleBytes~B), L1D capacity
(\MeasuredPmuLoneDCapKiB~KiB), and an empirical hierarchy curve, but it does not
recover a physical set-index hash. The 16-KiB page-stride conflict sweep crosses
the high-miss boundary at \MeasuredPmuCongruentKneeLow--\MeasuredPmuCongruentKneeHigh\
dependent lines depending on page offset, so a single associativity number is
not justified. Apple silicon exposes neither user-controllable physical
addresses nor a usable EL0 cycle counter in this configuration. A subtractive
PMU ``prime set plus one victim load'' experiment is included in the artifact;
on this machine its resident/DRAM calibration had no stable positive signal, so
we do not use it as evidence.

The unprivileged probe is retained alongside the privileged one to mark the
boundary: it records
\texttt{kpc\_set\_thread\_counting}~$=$~\MeasuredKpcSetFixedResult, whereas the
root path returns valid per-thread counts. Enabling a third-party kernel
extension or reducing SIP on Apple silicon requires RecoveryOS and a reboot; it
cannot be performed from this running macOS session. The artifact therefore
uses all measurements available to root in the normal OS and records where a
future RecoveryOS/kext path could add physical-address or timer control. The
GPU victim is exercised as a synthetic read-modify-write pass and as a
spatial-locality stencil, and the artifact includes the full randomized
resource-mode matrix; other production graphics kernels may still leave
different cache footprints.

\section{Reproducibility}

The arXiv ancillary directory contains all benchmark sources, pinned upstream
reference sources, build scripts, raw JSONL and upstream text/CSV output,
processed summaries, public IORegistry counter legends, plot tables,
environment capture, and SHA-256 checksums. The complete workflow is:

\begin{verbatim}
make
./scripts/run_experiments.sh
\end{verbatim}

Figures are generated by PGFPlots directly from processed data. Upstream
benchmark revisions, compiler version, array sizes, thread counts, block order,
trial counts, and the MCC/AGX IOReport channel inventory are recorded in
machine-readable files. The privileged PMU and IOReport measurements are
produced by \texttt{src/kpc\_counters.c} and \texttt{src/ioreport\_sampler.mm};
\texttt{run\_experiments.sh} invokes them automatically when sudo is available
and otherwise skips them, keeping the unprivileged timing run reproducible. The
source tree also includes the randomized resource-mode SLC probe matrix for
shared/private buffers and textures and the spatial-locality stencil victim.

\section{Conclusion}

M4 Pro retains GPU-induced cache state after a synchronized Metal command. A
16-MiB SLC-selective CPU traversal slows by \MeasuredSlcDeltaSixtyFourPct\% at a
64-MiB completed GPU footprint and by \MeasuredSlcDeltaFiveTwelvePct\% at
512 MiB, while a second traversal removes
\MeasuredSlcRecoveryFiveTwelvePct\% of the 512-MiB first-pass cost. The effect
persists across Metal buffer and texture storage modes, and Metal's texture
access-optimization calls do not erase it. Independent STREAM/BabelStream runs,
root PMU counters, and live IOReport agent histograms tie the timing result to
hardware behavior. The developer rule is concrete: phase synchronization
establishes correctness, not CPU cache residency; one planned CPU rewarm pass
can remove much of the measured transition penalty.

\bibliographystyle{unsrturl}
\IfFileExists{references.bib}{\bibliography{references}}{\bibliography{paper/references}}

\end{document}

%% file: results_values.tex

\newcommand{\MeasuredGpuUiSlowdownPct}{16}
\newcommand{\MeasuredGpuBgSlowdownPct}{-1}
\newcommand{\MeasuredPLoneDKiB}{128}

\newcommand{\MeasuredPLtwoMiB}{16}
\newcommand{\MeasuredELoneDKiB}{64}

\newcommand{\MeasuredELtwoMiB}{4}
\newcommand{\MeasuredCacheLineBytes}{128}
\newcommand{\MeasuredPageKiB}{16}
\newcommand{\MeasuredPCores}{10}
\newcommand{\MeasuredECores}{4}
\newcommand{\MeasuredPCpusPerLTwo}{5}
\newcommand{\MeasuredECpusPerLTwo}{4}
\newcommand{\MeasuredWarmUiOneMiBNs}{7.1}

\newcommand{\MeasuredWarmUiSixteenMiBNs}{80.3}
\newcommand{\MeasuredColdUiSixteenMiBNs}{94.3}

\newcommand{\MeasuredStreamTenGbps}{204.2}
\newcommand{\MeasuredBabelTenGbps}{199.4}
\newcommand{\MeasuredCustomTenGbps}{192.8}

\newcommand{\MeasuredCustomVsBabelAbsPct}{3}
\newcommand{\MeasuredSlcDeltaSixtyFourPct}{31}
\newcommand{\MeasuredSlcDeltaFiveTwelvePct}{63}

\newcommand{\MeasuredContiguousImmediateFiveTwelveNs}{2.65}

\newcommand{\MeasuredSlcImmediateSixtyFourNs}{4.10}
\newcommand{\MeasuredSlcRecoverySixtyFourNs}{3.14}
\newcommand{\MeasuredSlcRecoverySixtyFourPct}{24}
\newcommand{\MeasuredSlcImmediateFiveTwelveNs}{5.12}
\newcommand{\MeasuredSlcRecoveryFiveTwelveNs}{3.20}
\newcommand{\MeasuredSlcRecoveryFiveTwelvePct}{38}
\newcommand{\MeasuredCpuStressUiGbps}{189.8}
\newcommand{\MeasuredCpuStressBgGbps}{15.6}

\newcommand{\MeasuredKpcSetFixedResult}{-1}
\newcommand{\MeasuredMccPublicChannelCount}{191}
\newcommand{\MeasuredAgxPublicChannelCount}{295}

%% file: results_values_pmu.tex
\newcommand{\MeasuredPmuLoneDRefillGranuleBytes}{64}
\newcommand{\MeasuredPmuLoneDWithinRefills}{1.01}
\newcommand{\MeasuredPmuLoneDAcrossRefills}{2.01}
\newcommand{\MeasuredPmuLoneDCapKiB}{128}
\newcommand{\MeasuredPmuLoneDHitCyc}{4.1}
\newcommand{\MeasuredPmuLtwoHitCyc}{28}
\newcommand{\MeasuredPmuDramCyc}{536}

\newcommand{\MeasuredPmuSlcSatMiB}{32}

\newcommand{\MeasuredPmuCongruentKneeLow}{7}
\newcommand{\MeasuredPmuCongruentKneeHigh}{8}
\newcommand{\MeasuredPmuPCoreGhz}{4.3}
\newcommand{\MeasuredPmuECoreGhz}{1.1}
\newcommand{\MeasuredPmuPsetCreateKr}{-303}
\newcommand{\MeasuredPmuAffinityKr}{46}
\newcommand{\MeasuredPmuNproc}{14}

%% file: results_values_ioreport.tex
\newcommand{\MeasuredIoAmccIdleState}{17}
\newcommand{\MeasuredIoAmccCpuUiState}{224}
\newcommand{\MeasuredIoAmccCpuBgState}{32}
\newcommand{\MeasuredIoAmccGpuState}{200}
\newcommand{\MeasuredIoAgxGpuSaturation}{100}
\newcommand{\MeasuredIoPaccCpuUiSaturation}{100}
\newcommand{\MeasuredIoEaccCpuBgSaturation}{64}

%% file: results_values_resource.tex
\newcommand{\MeasuredResourceFiveTwelveMinPct}{55.7}
\newcommand{\MeasuredResourceFiveTwelveMaxPct}{59.0}
\newcommand{\MeasuredResourceFiveTwelveSpreadPct}{3.3}
\newcommand{\MeasuredResourceCpuOptContrastPct}{0.2}
\newcommand{\MeasuredResourceCpuOptContrastLowPct}{-0.8}
\newcommand{\MeasuredResourceCpuOptContrastHighPct}{1.9}
\newcommand{\MeasuredResourceGpuOptContrastPct}{-0.5}
\newcommand{\MeasuredResourceGpuOptContrastLowPct}{-3.5}
\newcommand{\MeasuredResourceGpuOptContrastHighPct}{3.8}

%% file: figures.tex
\definecolor{mblue}{RGB}{0,92,153}
\definecolor{mred}{RGB}{190,62,45}
\definecolor{mgreen}{RGB}{45,126,85}
\definecolor{mgray}{RGB}{95,99,104}

\pgfplotsset{
  m4axis/.style={
    width=0.82\textwidth,
    height=6.1cm,
    grid=major,
    grid style={black!10},
    axis line style={black!55},
    tick style={black!55},
    tick label style={font=\scriptsize},
    label style={font=\footnotesize},
    title style={font=\footnotesize},
    legend style={
      draw=none,
      fill=none,
      font=\scriptsize,
      cells={anchor=west},
      /tikz/every even column/.append style={column sep=5pt}
    },
    error bars/error bar style={line width=0.45pt},
    error bars/error mark options={rotate=90, mark size=1.5pt},
  }
}

\newcommand{\referencevalidationfigure}{%
\begin{figure}[!t]
\centering
\begin{tikzpicture}
\begin{axis}[
  m4axis,
  xlabel={OpenMP or worker threads},
  ylabel={Triad bandwidth (GB/s)},
  xmin=0.5,
  xmax=14.5,
  ymin=80,
  ymax=225,
  xtick={1,2,4,6,8,10,14},
  legend columns=2,
  legend style={at={(0.5,1.04)},anchor=south},
]
\addplot+[
  thick, mark=*, color=mblue,
  error bars/.cd, y dir=both, y explicit
] table[
  x=threads, y=stream_median,
  y error minus=stream_err_minus, y error plus=stream_err_plus
] {plotdata/reference_triad.dat};
\addlegendentry{STREAM 5.10}
\addplot+[
  thick, mark=square*, color=mgreen,
  error bars/.cd, y dir=both, y explicit
] table[
  x=threads, y=babelstream_median,
  y error minus=babelstream_err_minus,
  y error plus=babelstream_err_plus
] {plotdata/reference_triad.dat};
\addlegendentry{BabelStream 5.0}
\addplot+[
  thick, mark=triangle*, color=mred,
  error bars/.cd, y dir=both, y explicit
] table[
  x=threads, y=custom_median,
  y error minus=custom_err_minus, y error plus=custom_err_plus
] {plotdata/reference_triad.dat};
\addlegendentry{artifact triad}
\end{axis}
\end{tikzpicture}
\caption{Reference validation. Points are medians; error bars are bootstrap
95\% confidence intervals over seven independent upstream runs or nine artifact
trials.}
\label{fig:reference-validation}
\end{figure}
}

\newcommand{\cachelatencyfigure}{%
\begin{figure}[!t]
\centering
\begin{tikzpicture}
\begin{axis}[
  m4axis,
  xlabel={Working set (MiB)},
  ylabel={Dependent-load latency (ns)},
  xmode=log,
  log basis x=2,
  ymode=log,
  log basis y=10,
  xmin=0.0035,
  xmax=145,
  ymin=1.2,
  ymax=170,
  xtick={0.00390625,0.015625,0.0625,0.25,1,4,16,64,128},
  xticklabels={4K,16K,64K,256K,1,4,16,64,128},
  legend columns=2,
  legend style={at={(0.5,1.04)},anchor=south},
]
\addplot+[
  thick, mark=*, color=mblue,
  error bars/.cd, y dir=both, y explicit
] table[
  x=working_set_mib, y=user_interactive_median,
  y error minus=user_interactive_err_minus,
  y error plus=user_interactive_err_plus
] {plotdata/cpu_latency_warm.dat};
\addlegendentry{warm}
\addplot+[
  thick, mark=square*, color=mred,
  error bars/.cd, y dir=both, y explicit
] table[
  x=working_set_mib, y=user_interactive_median,
  y error minus=user_interactive_err_minus,
  y error plus=user_interactive_err_plus
] {plotdata/cpu_latency_flushed.dat};
\addlegendentry{\texttt{sys\_dcache\_flush}}
\draw[densely dashed, color=mgray]
  (axis cs:0.125,1.2) -- (axis cs:0.125,160);
\draw[densely dashed, color=mgray]
  (axis cs:16,1.2) -- (axis cs:16,160);
\end{axis}
\end{tikzpicture}
\caption{Cache-line-granular pointer chasing on the user-interactive path.
Each node occupies one 128-byte line; flushed trials invalidate the full range
before every measured traversal.}
\label{fig:cache-latency}
\end{figure}
}

\newcommand{\slcrecoveryfigure}{%
\begin{figure}[!t]
\centering
\begin{tikzpicture}
\begin{axis}[
  m4axis,
  height=6.3cm,
  xlabel={Completed GPU victim footprint (MiB)},
  ylabel={CPU probe time (ns/line)},
  xmode=log,
  log basis x=2,
  xmin=0.45,
  xmax=620,
  ymin=2.45,
  ymax=4.65,
  xtick={0.5,1,4,16,64,256,512},
  xticklabels={0,1,4,16,64,256,512},
  legend columns=2,
  legend style={at={(0.5,1.04)},anchor=south},
]
\addplot+[
  very thick, mark=*, color=mred,
  error bars/.cd, y dir=both, y explicit
] table[
  x=gpu_victim_mib_plot, y=immediate_median,
  y error minus=immediate_err_minus,
  y error plus=immediate_err_plus
] {plotdata/slc_gpu_recovery.dat};
\addlegendentry{first CPU pass}
\addplot+[
  very thick, mark=square*, color=mblue,
  error bars/.cd, y dir=both, y explicit
] table[
  x=gpu_victim_mib_plot, y=recovery_median,
  y error minus=recovery_err_minus,
  y error plus=recovery_err_plus
] {plotdata/slc_gpu_recovery.dat};
\addlegendentry{second CPU pass}
\draw[densely dashed, color=mgray]
  (axis cs:64,2.45) -- (axis cs:64,4.60);
\end{axis}
\end{tikzpicture}
\caption{Residual GPU cache state. The Metal victim completes before either
CPU measurement. The first 16-MiB SLC-selective traversal slows as GPU footprint
increases; one CPU traversal restores the second-pass cost to the no-victim
range. Error bars are bootstrap 95\% confidence intervals over 51 trials.}
\label{fig:slc-recovery}
\end{figure}
}

\newcommand{\slcpatternfigure}{%
\begin{figure}[!t]
\centering
\begin{tikzpicture}
\begin{axis}[
  m4axis,
  height=6.2cm,
  xlabel={Completed GPU victim footprint (MiB)},
  ylabel={First CPU pass (ns/line)},
  xmode=log,
  log basis x=2,
  xmin=0.45,
  xmax=620,
  ymin=0.45,
  ymax=5.25,
  xtick={0.5,1,4,16,64,256,512},
  xticklabels={0,1,4,16,64,256,512},
  legend columns=2,
  legend style={at={(0.5,1.04)},anchor=south},
]
\addplot+[
  very thick, mark=*, color=mred,
  error bars/.cd, y dir=both, y explicit
] table[
  x=gpu_victim_mib_plot, y=selective_median,
  y error minus=selective_err_minus,
  y error plus=selective_err_plus
] {plotdata/slc_probe_patterns.dat};
\addlegendentry{8192-byte stride}
\addplot+[
  very thick, mark=square*, color=mblue,
  error bars/.cd, y dir=both, y explicit
] table[
  x=gpu_victim_mib_plot, y=contiguous_median,
  y error minus=contiguous_err_minus,
  y error plus=contiguous_err_plus
] {plotdata/slc_probe_patterns.dat};
\addlegendentry{contiguous lines}
\draw[densely dashed, color=mgray]
  (axis cs:64,0.45) -- (axis cs:64,5.15);
\end{axis}
\end{tikzpicture}
\caption{Address-pattern separation after completed GPU work. Both probes touch
16 MiB, but the 8192-byte-stride probe spans approximately 1 GiB and avoids the
line-streaming behavior visible in the contiguous-line control.}
\label{fig:slc-patterns}
\end{figure}
}

\newcommand{\qoscontentionfigure}{%
\begin{figure}[!t]
\centering
\begin{tikzpicture}
\begin{groupplot}[
  group style={group size=2 by 1, horizontal sep=1.4cm},
  m4axis,
  width=0.45\textwidth,
  height=5.1cm,
  ybar,
  enlarge x limits=0.28,
]
\nextgroupplot[
  title={Metal linear copy},
  ylabel={GPU GB/s},
  ymin=80,
  ymax=160,
  xtick={1,2,3},
  xticklabels={baseline,CPU-UI,CPU-BG},
]
\addplot+[
  fill=mblue!35, draw=mblue,
  error bars/.cd, y dir=both, y explicit
] table[
  x=x, y=median, y error minus=err_minus, y error plus=err_plus
] {plotdata/gpu_contention.dat};

\nextgroupplot[
  title={Concurrent CPU triad},
  ylabel={CPU GB/s},
  ymin=0,
  ymax=215,
  xtick={1,2},
  xticklabels={UI,BG},
]
\addplot+[
  fill=mgreen!40, draw=mgreen,
  error bars/.cd, y dir=both, y explicit
] table[
  x=x, y=median, y error minus=err_minus, y error plus=err_plus
] {plotdata/cpu_contention_stress.dat};
\end{groupplot}
\end{tikzpicture}
\caption{Concurrent contention in seven rotated blocks with matched ten-thread
CPU workloads. GPU error bars summarize block medians; CPU error bars summarize
the seven stress runs.}
\label{fig:qos-contention}
\end{figure}
}

\newcommand{\pmusweepfigure}{%
\begin{figure}[!t]
\centering
\begin{tikzpicture}
\begin{axis}[
  m4axis,
  height=6.3cm,
  name=mainaxis,
  xlabel={Working set},
  ylabel={Cycles / access (PMU)},
  xmode=log, log basis x=2,
  ymode=log, log basis y=10,
  xmin=14, xmax=150000,
  ymin=3, ymax=700,
  xtick={16,64,256,1024,4096,16384,65536,131072},
  xticklabels={16K,64K,256K,1M,4M,16M,64M,128M},
  axis y line*=left,
]
\addplot+[very thick, mark=*, mark size=1.2pt, color=mblue]
  table[x=ws_kib, y=cyc_per_access] {plotdata/pmu_cache_sweep.dat};
\label{plot:cyc}
\draw[densely dashed, color=mgray] (axis cs:128,3) -- (axis cs:128,700);
\draw[densely dashed, color=mgray] (axis cs:16384,3) -- (axis cs:16384,700);
\node[font=\scriptsize, color=mgray, anchor=south, rotate=90]
  at (axis cs:128,10) {L1D 128\,KiB};
\node[font=\scriptsize, color=mgray, anchor=south, rotate=90]
  at (axis cs:16384,10) {L2 16\,MiB};
\end{axis}
\begin{axis}[
  m4axis,
  height=6.3cm,
  xmode=log, log basis x=2,
  xmin=14, xmax=150000,
  ymin=-30, ymax=1050,
  axis y line*=right,
  axis x line=none,
  ylabel={L1D misses / 1000 loads},
  ylabel style={color=mred},
  ytick={0,250,500,750,1000},
]
\addplot+[thick, mark=square*, mark size=1.0pt, color=mred]
  table[x=ws_kib, y=l1d_miss_per_1k] {plotdata/pmu_cache_sweep.dat};
\end{axis}
\end{tikzpicture}
\caption{Root-PMU working-set sweep on a performance core. Blue (left, log):
dependent-load latency in cycles, hardware-measured. Red (right): retired L1D
load misses per 1000 loads, which stay near zero through 128\,KiB and saturate
immediately above it, fixing the L1D capacity; the latency leaves its
$\sim$4-cycle L1D and $\sim$28-cycle L2 plateaus and climbs toward the DRAM
plateau ($\sim$510 cycles) by the tens of MiB. The PMU table-walk counter (not
shown) confirms TLB misses stay below 5\% of accesses up to 32\,MiB, so this
rise is system-level-cache residency rather than translation overhead.}
\label{fig:pmu-sweep}
\end{figure}
}

\newcommand{\pmugranulefigure}{%
\begin{figure}[!t]
\centering
\begin{tikzpicture}
\begin{axis}[
  m4axis,
  height=5.7cm,
  xlabel={Second dependent load offset (bytes)},
  ylabel={Events per load pair},
  xmin=-8,
  xmax=264,
  ymin=0.75,
  ymax=2.25,
  xtick={0,32,64,96,128,192,256},
  legend columns=2,
  legend style={at={(0.5,1.04)},anchor=south},
]
\addplot+[
  very thick, mark=*, color=mred,
  error bars/.cd, y dir=both, y explicit
] table[
  x=offset_bytes, y=refill_median,
  y error minus=refill_err_minus,
  y error plus=refill_err_plus
] {plotdata/pmu_l1d_granule.dat};
\addlegendentry{L1D refills}
\addplot+[
  thick, mark=square*, color=mblue,
  error bars/.cd, y dir=both, y explicit
] table[
  x=offset_bytes, y=miss_median,
  y error minus=miss_err_minus,
  y error plus=miss_err_plus
] {plotdata/pmu_l1d_granule.dat};
\addlegendentry{retired load misses}
\draw[densely dashed, color=mgray]
  (axis cs:64,0.75) -- (axis cs:64,2.2);
\end{axis}
\end{tikzpicture}
\caption{L1D refill granularity from dependent two-load pairs. Each random
record is flushed before the pass. A dependent second load within 32 bytes
shares one L1D refill with the first load; at a 64-byte offset the median jumps
to two refills.}
\label{fig:pmu-granule}
\end{figure}
}

\newcommand{\pmuassocfigure}{%
\begin{figure}[!t]
\centering
\begin{tikzpicture}
\begin{axis}[
  m4axis,
  width=0.78\textwidth,
  height=5.6cm,
  xlabel={Congruent lines in dependent cycle},
  ylabel={Page offset},
  xmin=0.5,
  xmax=20.5,
  ymin=0.5,
  ymax=5.5,
  xtick={1,4,8,12,16,20},
  ytick={1,2,3,4,5},
  yticklabels={0,64,128,4096,8192},
  colormap={m4heat}{color(0cm)=(mblue!20); color(1cm)=(mred!90)},
  colorbar,
  colorbar style={ylabel={$\log_{10}$(misses/1k+0.01)}, yticklabel style={font=\scriptsize}},
  scatter,
  only marks,
  mark=square*,
  mark size=5.5pt,
  point meta=explicit,
]
\addplot table[
  x=ways, y=offset_id, meta=log_miss
] {plotdata/pmu_l1d_assoc_heatmap.dat};
\end{axis}
\end{tikzpicture}
\caption{Page-offset-dependent congruent conflict sweep. The 16-KiB page stride
does not give a single clean associativity number on M4 Pro: different offsets
cross the high-miss boundary at different cycle lengths, showing that the set
function is not recoverable from page-offset bits alone.}
\label{fig:pmu-assoc}
\end{figure}
}

\newcommand{\resourcefigure}{%
\begin{figure}[!t]
\centering
\begin{tikzpicture}
\begin{groupplot}[
  group style={group size=2 by 1, horizontal sep=2.0cm},
  m4axis,
  width=0.38\textwidth,
  height=5.5cm,
]
\nextgroupplot[
  title={Resource-mode response},
  xlabel={GPU footprint},
  ylabel={Resource path},
  xmin=0.5,
  xmax=3.5,
  ymin=0.5,
  ymax=7.5,
  xtick={1,2,3},
  xticklabels={16,64,512},
  ytick={1,2,3,4,5,6,7},
  yticklabels={B-sh,B-pr,T-sh,T-cpu,T-gpu,T-linear,T-pr},
  colormap={m4heat}{color(0cm)=(mblue!20); color(1cm)=(mred!85)},
  colorbar,
  colorbar style={
    width=0.16cm,
    ylabel={\% above block baseline},
    ylabel style={font=\scriptsize},
    yticklabel style={font=\scriptsize}
  },
  scatter,
  only marks,
  mark=square*,
  mark size=6pt,
  point meta=explicit,
]
\addplot table[
  x=footprint_id, y=condition_id, meta=delta_pct
] {plotdata/slc_resource_heatmap.dat};
\nextgroupplot[
  title={512 MiB footprint},
  xlabel={\% above block baseline},
  ylabel={},
  xmin=45,
  xmax=70,
  ymin=0.5,
  ymax=7.5,
  ytick={1,2,3,4,5,6,7},
  yticklabel=\empty,
]
\addplot+[
  only marks, mark=*, color=mred,
  error bars/.cd, x dir=both, x explicit
] table[
  y=condition_id, x=delta_pct,
  x error minus=delta_err_minus,
  x error plus=delta_err_plus
] {plotdata/slc_resource_512.dat};
\end{groupplot}
\end{tikzpicture}
\caption{Randomized resource-mode matrix. B-sh/B-pr are shared/private buffers;
T-sh/T-pr are shared/private textures; T-cpu and T-gpu call Metal's texture
access optimizers; T-linear disables GPU-optimized texture contents. The
residual penalty clusters across paths, so descriptor choices do not flush the
post-GPU cache state.}
\label{fig:resource-matrix}
\end{figure}
}

\newcommand{\ioreportagentfigure}{%
\begin{figure}[!t]
\centering
\begin{tikzpicture}
\begin{axis}[
  m4axis,
  width=0.76\textwidth,
  height=5.8cm,
  xlabel={Condition},
  ylabel={DCS agent histogram},
  xmin=0.5,
  xmax=4.5,
  ymin=0.5,
  ymax=5.5,
  xtick={1,2,3,4},
  xticklabels={idle,CPU-UI,CPU-BG,GPU},
  ytick={1,2,3,4,5},
  yticklabels={AMCC,AGX,PACC0,PACC1,EACC0},
  colormap={m4agent}{color(0cm)=(mblue!18); color(1cm)=(mgreen!40); color(2cm)=(mred!90)},
  colorbar,
  colorbar style={ylabel={Normalized histogram state (\%)}, yticklabel style={font=\scriptsize}},
  scatter,
  only marks,
  mark=square*,
  mark size=7pt,
  point meta=explicit,
]
\addplot table[
  x=condition_id, y=agent_id, meta=normalized_median
] {plotdata/ioreport_agent_heatmap.dat};
\end{axis}
\end{tikzpicture}
\caption{Live IOReport DCS-agent separation in seven rotated blocks. The
high-priority CPU stress saturates the performance-core agents, background QoS
shifts demand toward the efficiency agent, and the GPU stress saturates the AGX
agent while all conditions are observed through the same memory-controller
histogram surface.}
\label{fig:ioreport-agents}
\end{figure}
}